# Thermal Conductivities and Mechanical Properties of Epoxy Resin as a Function of the Degree of Cross-linking

Xiao Wan[1,3#], Baris Demir[2,5#], Meng An[4], Tiffany R. Walsh[2]*, Nuo Yang[1,3]*

1. State Key Laboratory of Coal Combustion, Huazhong University of Science and Technology, Wuhan 430074, China.
2. Institute for Frontier Materials, Deakin University, Geelong, VIC 3216, Australia.
3. Nano Interface Center for Energy (NICE), School of Energy and Power Engineering, Huazhong University of Science and Technology, Wuhan 430074, China.
4. College of Mechanical and Electrical Engineering, Shaanxi University of Science and Technology, 6 Xuefuzhong Road, Weiyangdaxueyuan district of Xi'an, 710021, China.
5. Current address: Centre for Theoretical and Computational Molecular Science, The Australian Institute for Bioengineering and Nanotechnology, The University of Queensland 4072, Australia.

#X.W. and B.D. contributed equally to this work.
*Corresponding email: tiffany.walsh@deakin.edu.au (T.W.); nuo@hust.edu.cn (N.Y.)

# Abstract

Epoxy resins are widely used polymer matrices for numerous applications. Despite substantial advances, the molecular-level knowledge-base required to exploit these materials to their full potential remains limited. A deeper comprehension of structure/property relationships in epoxy resins at the molecular level is critical to progressing these efforts. It can be laborious, if not impractical, to elucidate these relationships based on experiments alone. Here, molecular dynamics simulations are used to calculate and compare thermal conductivities and mechanical properties of an exemplar epoxy resin, Bisphenol F cross-linked with Diethyl Toluene Diamine, revealing these inter-relationships. Both elastic modulus and thermal transport of the epoxy resin show an increase with greater cross-linking. Specifically, decomposition of the thermal conductivity into different force contributions suggests that the bonded term contributes to an increase in the heat flux. These outcomes provide a foundation for designing and fabricating customized epoxy resins with desirable thermal and mechanical attributes.

# 1. Introduction

Epoxy resins are important thermoset polymers widely employed in microelectronic industry as coatings, adhesives, [1, 2] encapsulants. [3] Due to their universal nature, the thermo-mechanical behavior of epoxy resins plays a vital role in the performance and reliability of electronics packaging, particularly from a perspective of thermal management. Furthermore, the thermo-mechanical properties of these materials depend significantly on the molecular-scale structure. To design electronics packaging and devices with reliable mechanical and thermal performance, it is important to understand the structure-property relationships of epoxy resins at the molecular scale. Advancement of this knowledge provides a solid foundation from which the properties of epoxy composites can be adapted via molecular design.

Over the last few years, many experimental researches have focused on the properties of epoxy-based composites and energy transport in these materials, which has enhanced our understanding of structure-property relationships for these materials. [4-11] The majority of these studies focused on their mechanical properties, whereas the number of studies about the thermal properties of epoxies is relatively few. [1, 3, 12-16] More specifically, where used to join two thermal conductive materials, for example in electronics packaging, thermal transport properties of epoxy resins gets quite crucial. [17] However, reported investigations on thermal transport properties of epoxy resins are relatively fewer in number. In an early study, Biercuk *et al.* studied the thermal conductivities of epoxy-based composites and found the measured thermal conductivity of neat epoxy resin, synthesized from diglycidyl ether of Bisphenol-F (EPON-862) and dimethane-amine curing agent, was ~0.2 W m$^{-1}$ K$^{-1}$ at 300 K. Moreover, it was slightly enhanced with an increase in temperature from 20 to 300 K. [18] Several studies since appeared that focused on thermal conductivity measurements of epoxy resins by different techniques, suggesting values in the range of 0.195-0.255 W m$^{-1}$ K$^{-1}$. [4, 5, 7, 19] In addition to thermal conductivity, the mechanical properties of epoxy resins are also of high interest. For example, Littell *et al.* investigated elastic modulus and Poisson's ratio of EPON-862 based epoxy resins in detail. [16] In this study Young's modulus was found to decrease with increasing temperature, whereas Poisson's ratio remained approximately constant, at 0.40-0.43 for different strain rates at room temperature. [16] There are also published works on the investigation of thermal properties of epoxies, such as glass transition temperature. [16, 20, 21] For instance, Miyagawa *et al.* found the glass transition of EPON-862 cured with methyltetrahydrophthalic-anhydride to be ~140 ℃. [21]

Although progress on the experimental evaluation of thermo-mechanical properties of epoxy is considerable, there are some limitations to consider in the experimental measurements. For example, the degree of cross-linking (DoC) is a crucial parameter affecting the thermo-mechanical properties of epoxy resins, which is the ratio of reacted epoxy carbon atoms to the total epoxy carbon atoms. [22, 23] Although there are approaches to infer the DoC such as swelling, tensile test, etc. these can only indirectly probe the degree of cross-linking, [24] and cannot yield immediate links between the molecular-scale structure of the materials and their thermal and mechanical properties. Besides, most experimental measurements are typically very time-consuming and expensive. [1, 12] Hence, a number of researchers have explored epoxy resins through computer simulation methods, such as molecular dynamic (MD) simulations. [1, 12-14, 25-31]

For the simulations of epoxy resins, first and foremost, the cross-linked network must be constructed carefully. Many researchers have contributed to the development of computational cross-linking methodologies. [12, 14, 15, 22, 26, 27] For instance, Yarovsky and Evans applied one-step reaction for the cross-linking procedure of epoxy resin called static cross-linking, in which all bonds between reactive atoms were generated synchronously. [15] This strategy is easily realized, but it cannot be used to control the degree of cross-linking in each epoxy resin system. Later developments, such as reported by Varshney *et al.*, introduced a dynamic cross-linking method, which applied a repeatedly cross-linking algorithm, resulting in a dynamical cross-linking process, along with multi-step relaxation procedure during new bond formation, to obtain the structure with different degrees of cross-linking. Furthermore, this approach can yield better equilibrated cross-linked structures with lower internal stresses. [26] The accumulated internal stresses in cross-linked systems may cause unrecoverable destruction on the structures when conducting simulations. [32] However, the success of this protocol also relies on a satisfactory equilibration the precursor liquid mixture sample prior to application of the cross-linking algorithm. In this vein，Demir and Walsh introduced a specific dynamic cross-linking process which is robust and repeatable. [22] Therefore, the cross-linking procedure introduced by Demir and Walsh was used in the current work presented herein. [22]

Given that a computationally-created cross-linked network of epoxy resin can be constructed with complete control over the network topology, many researchers have focused on thermal and mechanical properties of epoxy resins. [12, 13, 22, 25, 27, 33] In 2007, Fan *et al.* investigated the linear thermal expansion coefficient and Young's modulus of EPON-862 cross-linked with triethylenetetramine using molecular dynamics (MD) simulation. [27] These authors reported predicted properties that were consistent with experimental data, but did not provide sufficient technical detail to enable reproducibility. [27] Subsequently, Varshney *et al.* reported calculations of thermal conductivity of epoxy resin (EPON-862 cured with diethyltoluenediamine (DETDA)) via molecular dynamics (MD) simulations using their dynamic cross-linking method. [14] Their predicted thermal conductivity tallies well with the experimental values (0.2-0.3 W $m^{-1}$ $K^{-1}$). [14] Furthermore, Bandyopadhyay and Yang *et al.* predicted thermal expansion coefficients and elastic properties of EPON-862/DETDA epoxy resin with DoC values based on Yarovsky and Evan's approach. [1, 12] These authors found that the elastic properties increased with DoC, whereas on the contrary, the thermal expansion coefficient decreased with DoC. However, the influence of the presence of the C-N cross-link bonds on thermo-mechanical behavior, especially the thermal conductivity, in *dynamically* cross-linked epoxy networks has been relatively under-investigated. [14] The influence of DoC on a wide range of thermo-mechanical properties of dynamically cross-linked epoxy resins need to be studied in detail.

Here, to further explore the effect of DoC on thermo-mechanical properties, the dynamical cross-linking procedure was used to prepare epoxy resin samples based on EPON-862 and DETDA, [22] and calculations of thermo-mechanical properties have been conducted. Thereafter, bonded and non-bonded contributions towards the thermal conductivity were evaluated and compared. We also studied the effect of temperature on the thermal conductivity and have compared our simulation results with experimental findings.

## 2. Method

All simulations in this work were performed using the Large-scale Atomic/Molecular Massively Parallel Simulation (LAMMPS) package [2, 34-38] developed by Sandia National Laboratories (3 Mar 2020, https://lammps.sandia.gov). [39] The interatomic interactions were described by the DREIDING force-field, [40] which is intended for dynamics of polymers and organic materials. The cut-off distance of coulombic interactions was set to 12 Å. Periodic boundary conditions were applied in all three dimensions. The contribution of long-range interactions was calculated via the particle-particle-particle-mesh (PPPM) solver. [41]

The details of sample preparation are shown in **Supplemental Material (SM)**. The molecular structures of EPON and DETDA are provided in **Fig. 1a and 1b.** Specifically, based on the cross-linking process previously proposed by Demir and Walsh, [22] the cross-linking bonds (C-N) were formed between EPON-862 and DETDA molecules. These new C-N bonds led to the constitution of a 3-D polymer network, as shown in **Fig. 1c**. An example snapshot of the resultant cross-linked epoxy resin structure is given in **Fig. 1d**. As the current work chiefly focuses on thermo-mechanical properties, interested readers are referred to Ref. [22] for the detailed cross-linking procedure. We investigated three DoC values in this work; 25%, 50% and 78%. The upper value of 78% was chosen to allow for backward compatibility and comparison with previously-reported simulations of the EPON-DETDA resin.

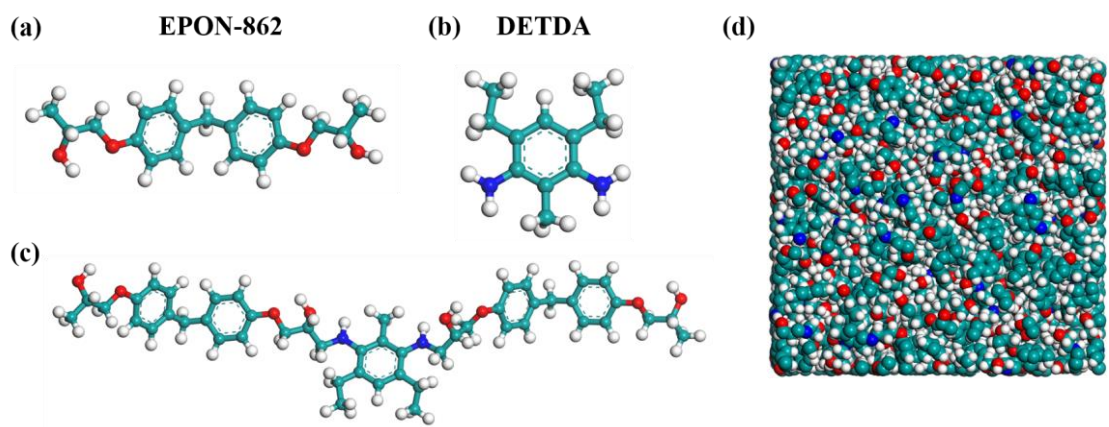

Fig. 1 (a, b) Molecular structures of monomer EPON-862 (the activated form) and curing agent DETDA. Color code: cyan, C; blue, N; red, O; white, H. (c) Product of reaction between two EPON molecules and the primary amine of the DETDA molecule. (d) Example snapshot of the relaxed cross-linked epoxy resin.

Having independently prepared three cross-linked epoxy resin samples with different DoC values, we then predicted a series of thermo-mechanical properties of each system for structure verification. Specifically, the glass transition temperature ($T_g$), coefficient of volumetric expansion ($\alpha_V$), elastic modulus, and Poisson's ratio are considered here. The detailed prediction of these properties can be found in **SM IV**.

In terms of thermal conductivities, non-equilibrium molecular dynamics (NEMD) simulations [39] were employed in the prediction of thermal conductivities of cross-linked epoxy resin samples, which has been widely used to get thermal transport properties. [42-46] The velocity Verlet algorithm was employed to integrate the equations of motion with a time step of 0.25 fs. [47] The heat source and heat sink were set as 320 K and 280 K, respectively, using Langevin thermostats, as shown in **Fig. 2 (a)**. Several layers of atoms at both ends of the simulation domain was fixed in space to prevent the heat flux across the periodic boundaries. Besides, the translational drift of the sample is blocked by the fixed atoms, which helps extract the temperature profile. [35]

The thermal conductivity of amorphous polymers is relatively small according to previous reports. [48] Hence, a temperature difference of 40 K is employed to get a measurable temperature gradient across the system. The system (except the fixed atoms) runs in the NVE ensemble for 1.5 ns to calculate thermal conductivity. The heat flux ($J$) is recorded by the average of the energy input and output rates from two baths per unit cross sectional area. On the basis of Fourier's law, the thermal conductivity ($\kappa$) was calculated as $\kappa = -J/\nabla T$, where $\nabla T$ is the temperature gradient across the sample. The temperature gradient was obtained by linear fitting to the temperature profile, as shown in **Fig. 2 (b)**. In each independent simulation, data at six different time intervals (0.25 ns each) in the steady state were used for thermal conductivity calculation. The error bars are the standard deviations of these data.

NEMD simulations were also used to study the effect of cross-link bonds on the resulting calculated heat flux. Utilizing LAMMPS software, [39] the total heat flux can be decomposed into respective interaction contributions from convection, non-bonded, bond, angle, and additionally, the dihedral term, while the sum of contributions from last three terms corresponds to bonded contributions in this work, so can the thermal conductivity. [2] Actually, the convection contribution term can be negligible in solid-like polymers due to restricted mobility of molecules in the samples. [2] For this reason, we didn't take convection term into account in the analysis for thermal transport properties of epoxy resins below. It should be emphasized that heat flux calculation in MDs by LAMMPS suffers from error for angle, dihedral and improper contributions. And this work only focuses on the impact of new cross-linking bonds on heat transfer and the contrast between non-bonded and bonded contributions in the epoxy resins with different degree of cross-linking (DoC), instead of the individual contributions of bond, angle, and dihedral terms. Hence, bonded contributions can be obtained by subtracting non-bonded contributions from the energy flow rate from the heat bath per unit area. The heat flux decomposition details are given in **SM V**.

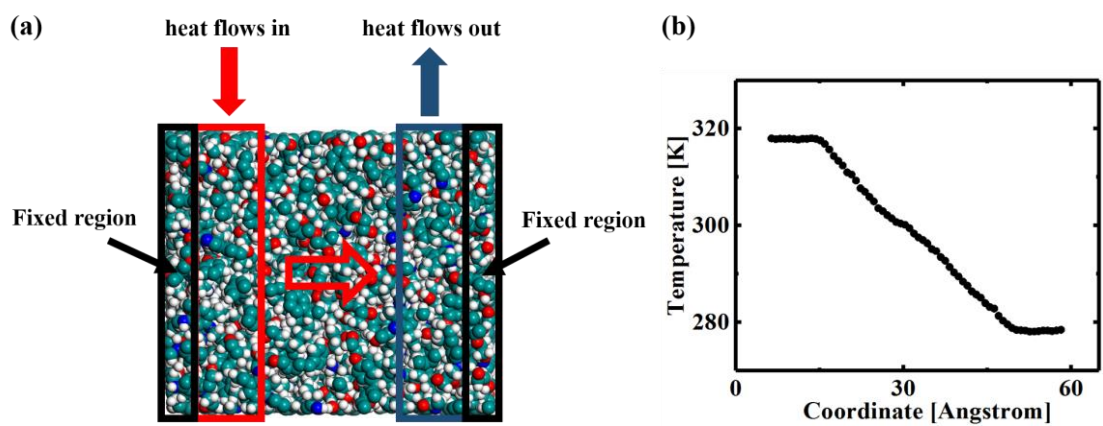

Fig. 2 (a) Simulation setup for thermal conductivity calculations in the non-equilibrium molecular dynamics simulations: heat flows across the sample from heat source to heat sink with the fixed regions setting at the end of the system. (b) Temperature profile of epoxy resin, used for calculating the thermal conductivity.

## 3. Results and discussion

Upon on the completion of the sample preparation, the polymerized samples were used to predict the thermo-mechanical properties. We then compare our predicted values with the experimentally available values reported in the literature.

### 3.1. Mechanical properties

Density versus temperature curves are shown in **Fig. 3 (a)** for the 25%, 50% and 78% cross-linked systems. Predicted density values over three samples are obtained as 1.130-1.180 g cm$^{-3}$ at room temperature, which are in good agreement with the experimental values ~1.200 g cm$^{-3}$. [22] As expected, the system with a higher DoC has a higher density at a given temperature, due to the greater number of cross-link bonds, which leads to a smaller volume.

Linear regression was used to fit the volumetric expansion plots shown in **Fig. 3 (b)** to calculate the coefficient of volumetric thermal expansion (CVTE) in both the rubbery and glassy states for three cross-linked systems. The rubbery and glassy regimes were identified from calculations of the glass transition temperature ($T_g$). From **Fig. 3 (c)**, it is clear that $T_g$ increases with an increasing degree of cross-linking. We found that $T_g$ increased from 382 K for the sample with a DoC of 25% to 443 K for the sample with a DoC of 78%. As mentioned above, the denser epoxy resin samples associated with the greater cross-linking degree led to a higher glass transition temperature. There is more resistance to increases in free volume as the temperature increases for the higher degrees of cross-linking. [12] The fact that the (raw) predicted $T_g$ is higher than published experimental data is an artefact of the cooling rate in MD simulations, [1] which is faster than experimental cooling rates. This leads to an over-estimate in predicted $T_g$ values here. In order to address this mismatch, the common correction is to add approximately 3 K per order of magnitude in the cooling rate relative to the experimental cooling rate. [49] Therefore, the adjusted simulation-based $T_g$ value should be approximately 30 K lower than the raw values.

The CVTE as a function of DoC is provided in **Fig. 3 (d)**. The rubbery and glassy CVTE both decrease as the DoC increases, which is likely due to higher degree of constraint of the polymer chains in the higher-DoC systems, resulting in less mobility during the thermal expansion process. The decreasing trend shows good consistency with previous studies, [12] with our results consistent with experimentally-measured values $\alpha_{g,exp} = $ 195 ppm K$^{-1}$ and $\alpha_{r,exp} = $ 579 ppm K$^{-1}$, respectively. [1, 12, 49, 50]

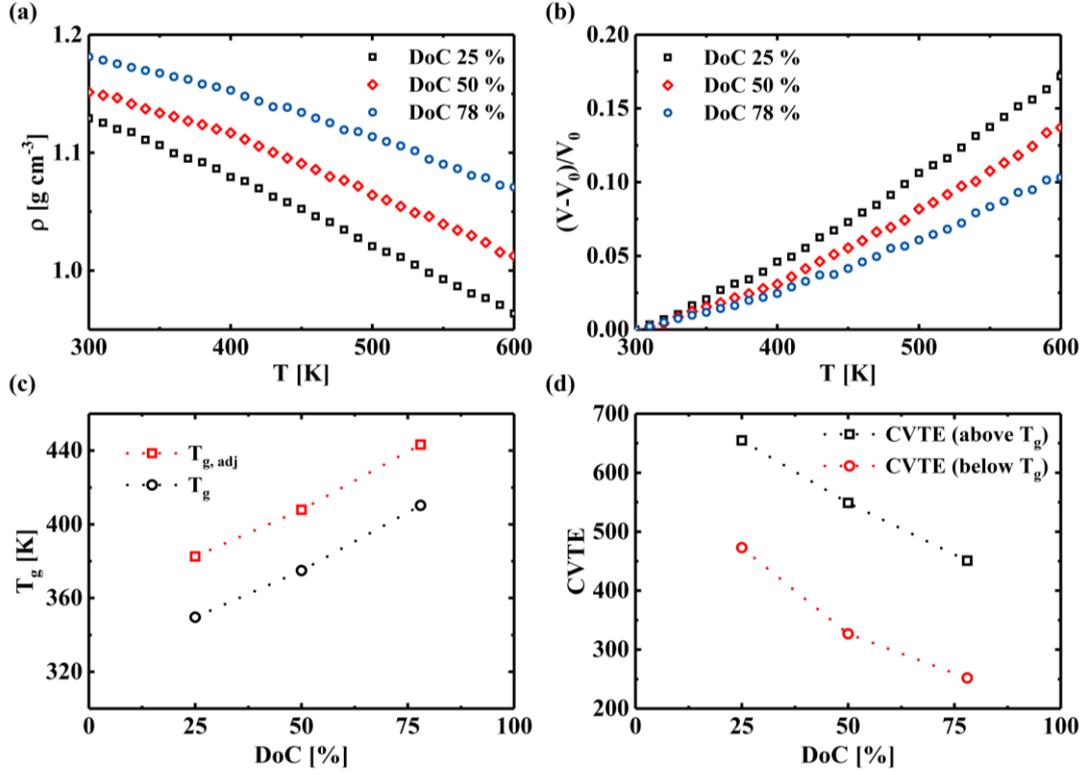

Fig. 3 Predicted thermal properties for resins with 25%, 50% and 78% degrees of cross-linking. (a) Density vs. temperature curves, (b) Volume dilation with respect to temperature, (c) the glass transition temperature $T_g$ (in both the raw and adjusted forms), and (d) the coefficient of volumetric thermal expansion (CVTE).

Elastic properties of the epoxy resin are considered next. **Fig. 4** provides the SSCs of the three cross-linked samples at 300 K, indicating that Young's modulus and the shear modulus increase with the degree of cross-linking. The values of these moduli and Poisson's ratio for the different cross-linked structures are summarized in **Table 1**. Our results are comparable to the previous experiment and simulation values. [16, 27] The discrepancy between predicted results and experimental values of these moduli can be attributed to higher strain rate. It is reported that the shear modulus of EPON-862 increased with an increase in the shear rate. [51] Hence, the cross-linking degree dependence of modulus is chiefly focused on here. Specifically, we found a stiffening of the epoxy resin from 1.80 GPa to 3.01 GPa as the DoC increased from 25% to 78%, again attributed to the presence of a greater number of cross-link bonds. We emphasize here that the importance of these results lies in the model validation and predicted trends as a function of DoC.

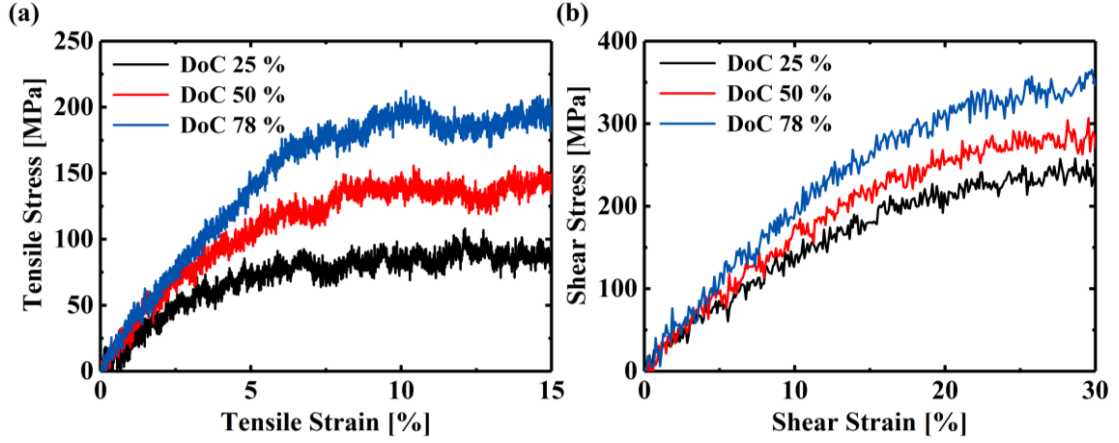

Fig. 4 (a) Tensile stress-strain curves for 25%, 50% and 78% cross-linked system. (b) Shear stress-strain curves for 25%, 50% and 78% cross-linked system.

**Table 1.** Predicted elastic properties (units in GPa) of the epoxy as a function of degree of cross-linking and compared with experimental values.

| DoC (%) | E (Young's modulus) | G (Shear modulus) | $\nu$ (Poisson's ratio) |
| --- | --- | --- | --- |
| 25 | 1.80 | 1.83 | 0.45 |
| 50 | 2.58 | 2.03 | 0.38 |
| 78 | 3.01 | 2.32 | 0.38 |
| Experimental sample [12, 16] | 1.60-2.90 | 0.60-1.00 | 0.35-0.43 |

## 3.2. Thermal conductivity

Temperature profile across the sample is shown in **Fig. 2 (b)**, which exhibits a good linear distribution in the simulation system except two heat bath regions. The thermal conductivities predicted for the epoxy resin samples as a function of temperature are provided in **Fig. 5**. Although experimentally-produced epoxy resin samples can feature a wide variety of forms and properties depending on its production and process history, the experimentally-determined thermal conductivity of several epoxies (including

EPON-862/DETDA and EPON-862/dimethane-amine, etc.) lies typically in the range of ~0.2-0.3 W m$^{-1}$ K$^{-1}$. [18, 19, 23, 52-54] Therefore, there is reasonable agreement between the experimental and simulation data. Within error, the system temperature had no significant effect on the predicted thermal conductivity. On the contrary, the thermal conductivity was found to be clearly DoC dependent, for the temperature range studied here.

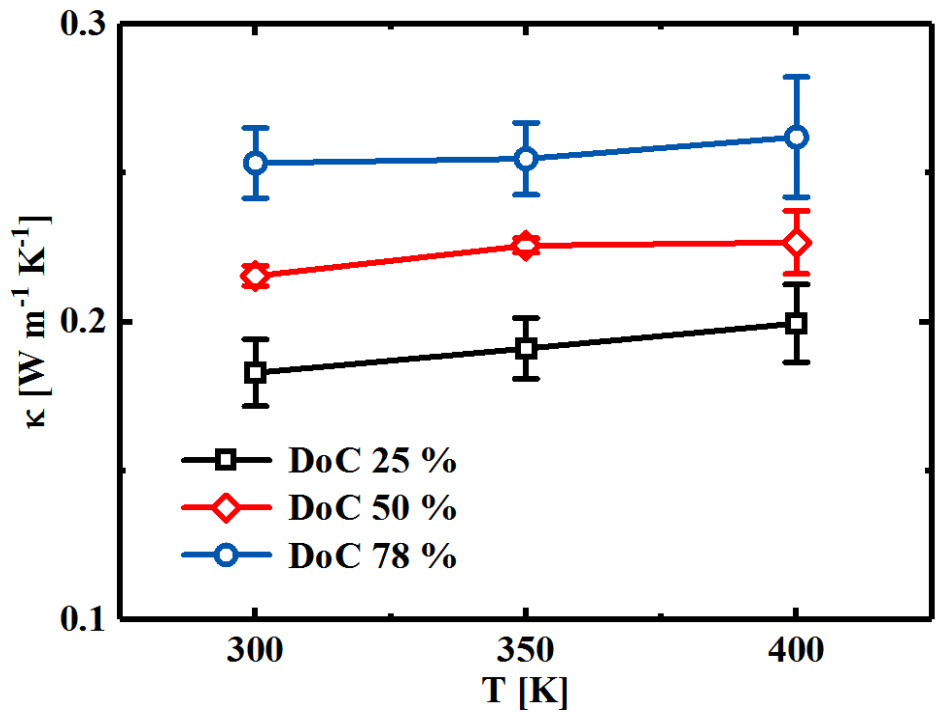

Fig. 5 The thermal conductivity (κ) vs. temperature curves for resins with 25%, 50% and 78% degrees of cross-linking.

In the cross-linked epoxy resin systems, the thermal conductivity at 300 K broadly increased with greater cross-linking degree, as indicated in **Fig. 6 (a)**. This tendency was consistently recovered in each of the three principal directions, which also indicates the isotropic character of our samples. To explain the DoC-dependent thermal conductivity in the epoxy resin systems, microscopic heat transfer mode was evaluated for different cross-linking degree samples. The decomposition of the thermal conductivity into contributions for our samples is provided in **Fig. 6 (b)**. The contributions are separated into two categories: (i) the energy transfer by bonded interactions (such as the bond stretching, angle bending and dihedral terms), and (ii) the energy transfer by non-bonded interactions including the contribution of intermolecular vdW and Coulomb interactions, labeled as "bonded" and "non-bonded", respectively. This decomposition identified that the non-bonded contribution was relatively weaker with an increasing degree of cross-linking. Furthermore, our data also indicate that the bonded contribution dominates for all degrees of cross-linking studied here.

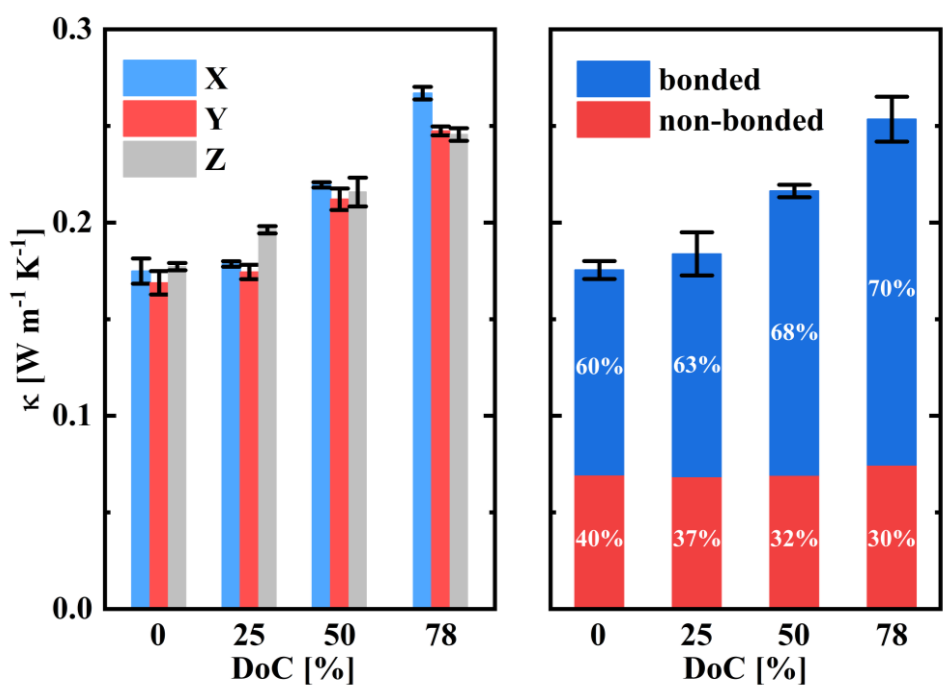

Fig. 6 (a) The thermal conductivity (κ) of the samples with 25%, 50% and 78% degree of cross-linking along each of the three principal (*x, y* and *z*) axes of simulation cell at 300 K. (b) Contributions of energy transfer modes to thermal conductivity in the resins as a function of the degree of cross-linking.

Besides, the bonded contribution increased with an increasing degree of cross-linking. This means that the introduction of more cross-linking bonds opens new pathways for heat transfer and directly leads to the enhancement of thermal transport.

Hence, it is concluded that the higher thermal conductivity in greater cross-linking degree epoxy resin can be attributed to the bonded interactions arising from the cross-linking bonds. The findings are consistent with previous simulation researches on amorphous polymers. [2, 55]

## Conclusions

In summary, we used molecular dynamics simulations to computationally-generate the structures and thermo-mechanical properties of dynamically cross-linked epoxy resins with different compositional parameters to investigate the structure/property relationships in these materials. The resulting predicted properties were generally aligned with reported experimental data. Our findings reveal the effect of cross-linking bonds on the thermo-mechanical performance of these materials. We found that elastic stiffness was enhanced by generation of new cross-linking bonds. The thermal conductivity was also found to broadly increase with the degree of cross-linking and was dominated by the contribution of the bonded interactions to the total heat flux. Our analysis suggested that the introduction of cross-linking bonds opens new pathways for thermal transport and results in an increase in thermal conductivity. Our simulation procedure and results provide a rational knowledge-base for designing and testing customized epoxy resins with desirable thermo-mechanical properties.

# Conflicts of interest

There are no conflicts of interest to declare.

# Acknowledgements

The work was sponsored by National Key Research and Development Project of China No. 2018YFE0127800 (N.Y.), Fundamental Research Funds for the Central Universities No. 2019kfyRCPY045 (N.Y.) and Program for HUST Academic Frontier Youth Team. We are grateful to Xiaoxiang Yu and Han Meng for useful discussions. The authors thank the National Supercomputing Center in Tianjin (NSCC-TJ) and the China Scientific Computing Grid (ScGrid) for providing assistance in computations. BD and TRW thank the National Computing Infrastructure and the Pawsey Supercomputing Centre, Australia, for access to computing resources under the NCMAS scheme. This work was partially supported by the Australian Research Council, grant number DP180100094.